\documentclass[conference]{IEEEtran}
\IEEEoverridecommandlockouts
\usepackage{cite}
\usepackage{amsmath,amssymb,amsfonts}
\usepackage{algorithmic}
\usepackage{graphicx}
\usepackage{textcomp}
\usepackage{xcolor}
\usepackage[version=4]{mhchem}
\usepackage{multirow}
\def\BibTeX{{\rm B\kern-.05em{\sc i\kern-.025em b}\kern-.08em
    T\kern-.1667em\lower.7ex\hbox{E}\kern-.125emX}}
\usepackage{array}
\usepackage[caption=false,font=footnotesize]{subfig}
\usepackage{url}

\begin{document}

\title{
A Gate-Based Quantum Computing Framework for Codon Optimization
}

\author{\IEEEauthorblockN{1\textsuperscript{st} Fatemeh Ghasemi}
\IEEEauthorblockA{\textit{Quantum Business Frontier Division (QBFD)} \\
\textit{Yonsei University}\\
International Campus, South Korea \\
fatemeh\_ghasemi@yonsei.ac.kr}
\and
\IEEEauthorblockN{2\textsuperscript{rd} Kion Kim}
\IEEEauthorblockA{\textit{Quantum Business Frontier Division (QBFD)} \\
\textit{Yonsei University}\\
International Campus, South Korea\\
kion.kim@yonsei.ac.kr
}}
\maketitle

\begin{abstract}

Codon optimization is a challenging combinatorial optimization problem with important applications in synthetic biology, protein expression, and biotechnology. While quantum annealing has previously been explored for this problem, the application of gate-based quantum algorithms remains largely unexplored. In this work, we present a gate-based quantum computing framework for codon optimization by formulating the optimization objective as an Ising Hamiltonian and expressing it in terms of Pauli operators suitable for gate-based quantum processors. The proposed framework is investigated using the Variational Quantum Eigensolver (VQE), Sampling Variational Quantum Eigensolver (SVQE), and the Quantum Approximate Optimization Algorithm (QAOA), with benchmarking against exact diagonalization, classical optimization, and quantum annealing. Using a three-amino-acid benchmark, the proposed implementations successfully reproduce the exact ground-state solutions while identifying optimized ansatz selections, optimization strategies, initialization schemes, and hardware execution settings that provide reproducible performance on current noisy quantum hardware. Hardware demonstrations are performed on IBM Quantum processors to evaluate the practical implementation of the proposed workflow. As a work in progress, this study establishes a reproducible computational framework for gate-based quantum codon optimization and provides the foundation for future investigations of larger protein sequences as quantum hardware and algorithms continue to advance.

\end{abstract}


\section{Introduction}

The genetic code exhibits a fundamental redundancy in which most amino acids are encoded by multiple synonymous codons. Although these codons correspond to the same amino-acid sequence, their usage can significantly affect protein expression, translation kinetics, mRNA stability, protein folding pathways, and other biological processes. Consequently, the design of coding sequences with desirable biological properties has become a central task in synthetic biology, metabolic and gene engineering, vaccine development, and recombinant protein production \cite{Gusta, Quax, Mauro}.
\begin{figure}
    \centering
    \includegraphics[width=0.9\linewidth]{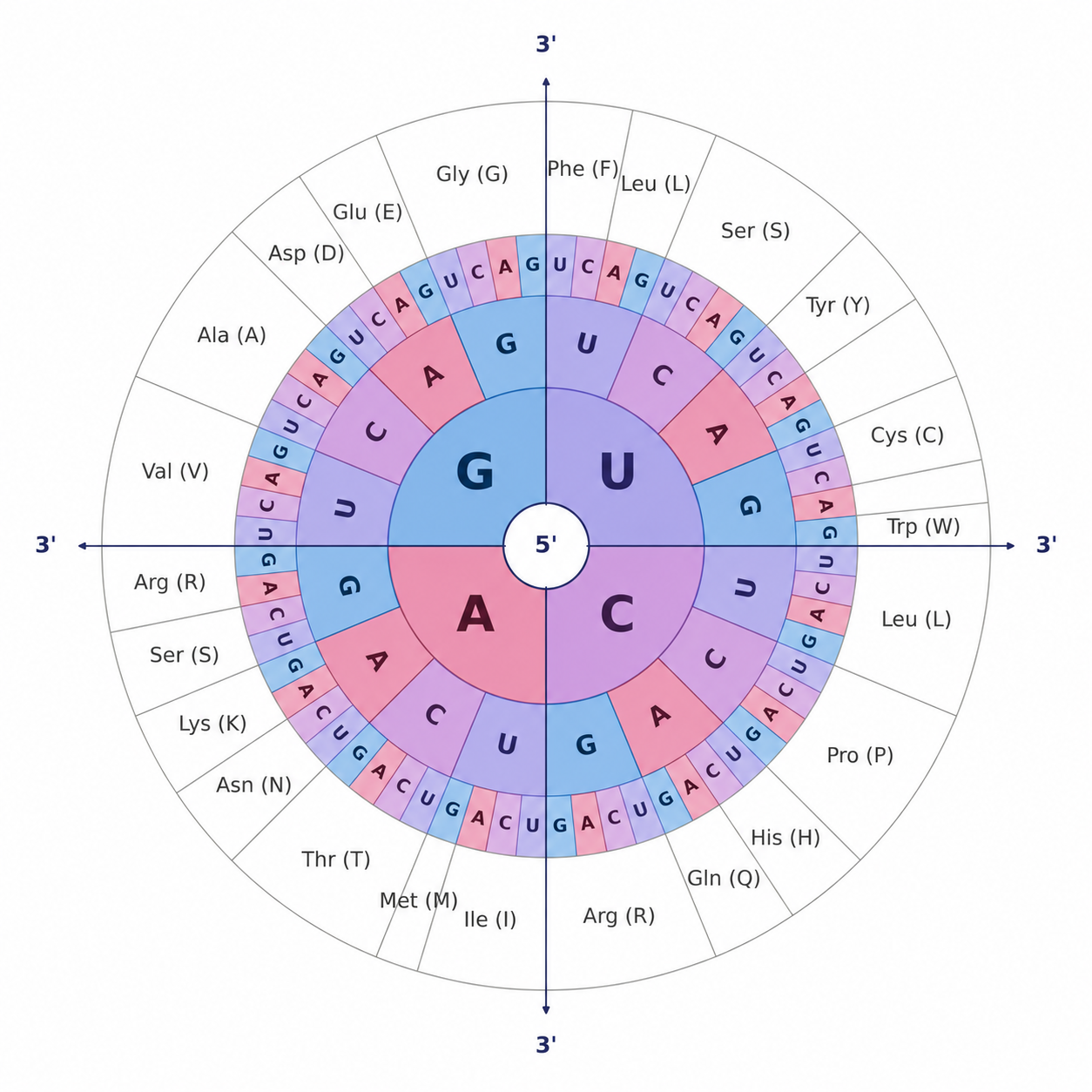}
    \caption{The genetic code wheel illustrating the mapping between mRNA codons and amino acids \cite{Gros}. Codons are read from the center \(5^{\prime}\) outward \(3^{\prime}\), corresponding to the first, second, and third nucleotide positions. The figure highlights the degeneracy of the genetic code, in which multiple synonymous codons encode the same amino acid. }
    \label{codon synonym}
    \vspace{-10pt}
\end{figure}
% \begin{figure}
%     \centering
%     \includegraphics[width=1\linewidth]{codon_wheel (3).png}
%     \caption{The genetic code wheel illustrating the mapping between mRNA codons and amino acids \cite{Gros}. Codons are read from the center (5) outward (3), corresponding to the first, second, and third nucleotide positions. The figure highlights the degeneracy of the genetic code, in which multiple synonymous codons encode the same amino acid. The green marker indicates the start codon (AUG), which initiates protein translation and encodes methionine (Met), whereas the red markers indicate the stop codons (UAA, UAG, and UGA), which terminate translation without encoding an amino acid.}
%     \label{codon synonym}
% \end{figure}

Most amino acids are encoded by more than one synonymous codon, meaning that many different nucleotide sequences can produce the same protein \cite{Fatima}. Fig. \ref{codon synonym} presents the synonymous codons corresponding to each amino acid. Codon optimization exploits this redundancy by selecting the most suitable synonymous codons for a target host organism while preserving the encoded amino acid sequence. The objective is to improve gene expression and protein production while satisfying some biological constraints and other sequence-dependent properties. Common optimization criteria include codon usage bias, codon-pair preferences, GC-content regulation, sequence repetitiveness, and organism-specific translational characteristics \cite{Gusta,Quax,Plotkin}. Because each amino acid may be represented by between one and six synonymous codons, the number of possible coding sequences increases exponentially with protein length. For a protein containing N amino acids, the search space ranges from $1^N$ to $6^N$ possible coding sequences. As a concrete example, a protein of only $100$ amino acids may have more than $10^{50}$
 possible synonymous coding sequences. Consequently, codon optimization becomes a large-scale combinatorial optimization problem for which exhaustive search is computationally infeasible.

Beyond its biological importance, codon optimization represents an attractive benchmark problem for quantum optimization algorithms. The problem naturally combines discrete binary decision variables, one-hot selection constraints, and competing objective functions within a highly structured combinatorial search space. These characteristics are shared by many practically relevant optimization problems while maintaining a direct connection to a real-world biological application \cite{Sen, Arbib}. %Furthermore, the availability of explicit Hamiltonian formulations enables direct implementation on both quantum annealing and gate-based quantum computing platforms.

Recent advances in quantum computing have stimulated considerable interest in applying quantum algorithms to combinatorial optimization problems \cite{Mont, Bhar}. Codon optimization represents a promising application for future quantum computers, which are expected to tackle such computationally demanding problems more efficiently than classical computers. 

Quantum annealing has emerged as a promising approach for codon optimization by mapping the problem onto a QUBO or constrained quadratic optimization model \cite{Fox}. 
A recent study introduced a biologically motivated QUBO formulation for codon optimization that incorporates codon-selection constraints together with several biologically relevant objective functions. This study successfully optimized biologically relevant genes, including the SARS-CoV-2 spike protein and human insulin, using D-Wave's hybrid constrained quadratic model (CQM) solver rather than the annealer alone \cite{Chung}.

%This study successfully optimized biologically relevant genes, including the SARS-CoV-2 spike protein and human insulin, using D-Wave's hybrid constrained quadratic model (CQM) solver rather than the annealer alone \cite{Chung}.

%Although quantum annealing provides a natural framework for solving the QUBO formulation of the codon optimization problem, it has several limitations. As a heuristic optimization method, it does not guarantee convergence to the global optimum. Moreover, the biological objectives and constraints must be encoded as QUBO penalty terms, which may introduce numerous auxiliary variables and require careful tuning of penalty coefficients. 

Although quantum annealing provides a natural framework for solving QUBO formulations, its practical implementation is limited by the need to encode all objectives and constraints as penalty terms, sparse qubit connectivity, embedding overhead, and analog hardware noise. The performance of current quantum annealers is constrained by these factors, which can all degrade the quality of the solution \cite{Yark, Katz, Pear, Konz}. Consequently, most practical implementations rely on hybrid quantum-classical solvers.

%, and the scalability of standalone quantum annealers for large and highly constrained codon optimization problems remains limited.

In contrast, gate-based quantum algorithms offer greater flexibility through programmable circuit architectures, variational ansatz, initialization strategies, classical optimization techniques, and advanced error-mitigation methods, making them an attractive platform for investigating increasingly complex codon optimization problems \cite{Fati}. Unlike quantum annealers, which are primarily designed for QUBO optimization, gate-based quantum computers provide a universal computational framework capable of implementing a wide range of quantum optimization algorithms. Biological constraints can be incorporated directly into the Hamiltonian or quantum circuit without requiring all objectives to be converted into QUBO penalty terms, reducing the need for auxiliary variables and simplifying problem formulation. Furthermore, gate-based architectures are compatible with quantum error correction, providing a scalable path toward fault-tolerant quantum computation. With the rapid growth of gate-based quantum optimization algorithms, benchmarking frameworks, and the increasing availability of cloud-accessible quantum processors, these platforms are expected to become increasingly suitable for solving large-scale combinatorial optimization problems, including codon optimization \cite{Koch}. These advantages motivate the development of gate-based quantum algorithms for codon optimization, with the aim of exploiting the flexibility, scalability, and long-term potential of universal quantum computers.
%while overcoming several limitations of current quantum annealing approaches.
The resulting formulation enables implementation on quantum optimization platforms and provides a useful foundation for systematic investigations of quantum algorithms in sequence design. 

In the present work, we employ the same objective function and biological constraints as the quantum annealing approach in Ref. \cite{Chung} to ensure a fair comparison with existing quantum annealing approaches. The Hamiltonian is represented in terms of Pauli operators, making it compatible with gate-based quantum computation. This construction enables the application of several quantum algorithms within a unified Hamiltonian framework. Here, we investigate the Variational Quantum Eigensolver (VQE), Sampling Variational Quantum Eigensolver (SVQE), and Quantum Approximate Optimization Algorithm (QAOA) for codon optimization. 
%VQE is a hybrid quantum-classical method that minimizes the expectation value of a problem Hamiltonian by iteratively optimizing the parameters of a variational quantum circuit. SVQE extends this approach by exploiting the probability distribution of the optimized quantum state to sample multiple candidate solutions.
%, making it particularly suitable for combinatorial optimization problems with degenerate or near-degenerate optima. 
%QAOA is another hybrid variational algorithm that alternates between problem-specific and mixing Hamiltonians to prepare low-energy solutions of optimization problems. 
All three methods employ parameterized quantum circuits whose variational parameters are optimized by a classical optimizer. They differ primarily in their ansatz construction, optimization strategy, and the way quantum measurement outcomes are utilized to identify optimal solutions. We also employ quantum annealing solvers, following the same implementation as in Ref. \cite{Chung}, as a benchmark for performance comparison.

To validate the proposed framework, we first consider a benchmark codon optimization problem based on the three-amino-acid sequence glutamic acid (Glu), glycine (Gly), and proline (Pro). Glu is encoded by two synonymous codons, Gly by four, and Pro by four, yielding ten candidate codons in total. Under the one-hot encoding scheme, each candidate codon is represented by a single binary variable, resulting in a 10-qubit optimization problem.
% four, yielding ten candidate codons in total and, consequently, a 10-qubit optimization problem.
The small problem size enables direct comparison with exact diagonalization while providing a well-defined test case for evaluating the performance of different quantum algorithms.
%Small benchmark systems consisting of short amino-acid sequences are first employed to validate the quantum optimization workflows against exact solutions. By examining multiple quantum algorithms across increasing problem sizes, we assess their applicability, scalability, and solution quality for codon optimization Hamiltonians.
The primary objective of this work is not to establish the superiority of a particular quantum algorithm, but rather to provide a systematic evaluation of several quantum optimization paradigms within a common biologically motivated framework. Such comparisons are important for understanding the strengths and limitations of current quantum-computing approaches and for identifying promising directions toward larger-scale quantum optimization in biological sequence design.

% ======================================================
\section{Hamiltonian Construction}
% ======================================================

The codon optimization problem is formulated as a binary optimization problem in which each binary variable represents the selection of a synonymous codon for a particular amino acid. For a protein sequence containing $L$ amino acids, let $q_i\in\{0,1\}$ denote the binary decision variable associated with codon $i$. A value of $q_i=1$ indicates that the corresponding codon is selected in a candidate coding sequence (bitstring), whereas $q_i=0$ indicates that it is not selected. Since each amino acid must be encoded by exactly one synonymous codon, the optimization is subject to a one-hot selection constraint,
$\sum_{i\in\mathcal{S}_a} q_i = 1$,
where $\mathcal{S}_a$ denotes the set of synonymous codons corresponding to amino acid $a$.

The total optimization Hamiltonian is constructed as

\begin{equation}
H = H_f + H_{GC} + H_R + H_P,
\label{eq:Htotal}
\end{equation}

where each term represents a biological objective or constraint. The codon usage term, $H_f$, promotes codons that are more frequently used by the target host organism. The GC-content term, $H_{GC}$, regulates the overall GC composition of the optimized nucleotide sequence toward a desired target value. The repeated-nucleotide term, $H_R$, penalizes long runs of identical nucleotides between neighboring codons to reduce undesirable sequence patterns. Finally, the penalty term, $H_P$, enforces the one-hot constraint, ensuring that exactly one synonymous codon is selected for each amino acid. The codon usage objective is defined as

\begin{equation}
H_f=-c_f\sum_i \log(F_i)\,q_i,
\label{eq:Hf}
\end{equation}

where $F_i$ denotes the usage frequency of codon $i$ in the target host organism and $c_f$ is the corresponding weighting coefficient. Consequently, codons with higher usage frequencies are assigned lower energies and are therefore favored during the optimization process. GC content is the fraction of guanine (G) and cytosine (C) nucleotides in a coding sequence. Maintaining an appropriate GC content is important for sequence stability, mRNA structure, and efficient gene expression in the target host. 
The GC-content objective is defined as
\begin{equation}
H_{\mathrm{GC}}
=
c_{\mathrm{GC}}
\left(
G-G_{\mathrm{target}}
\right)^2,
\end{equation}
where $G$ is the GC content of the candidate coding sequence, $G_{\mathrm{target}}$ is the desired GC content, and $c_{\mathrm{GC}}$ is the weighting coefficient. This quadratic penalty encourages the optimized sequence to maintain a GC content close to the specified target value.
%Accordingly, the GC-content objective is constructed by penalizing deviations of the overall GC fraction from a prescribed target value. Expanding this objective produces both linear and quadratic contributions, allowing the optimization to regulate the global nucleotide composition of the optimized sequence.
The repeated-nucleotide objective is defined as
\begin{equation}
H_R=c_R\sum_{k}R_k,
\end{equation}
where $R_k$ denotes the penalty associated with an undesirable repeated nucleotide pattern at position $k$, and $c_R$ is the corresponding weighting coefficient. This term penalizes consecutive repeated nucleotides, thereby promoting more balanced nucleotide distributions and reducing undesirable sequence patterns. The one-hot selection constraint is defined as

\begin{equation}
H_P = -\epsilon \sum_i q_i + \mathrm{infty}\sum_{a}
\sum_{\substack{i,j\in a\\ i<j}}q_iq_j,
\end{equation}

where the first summation rewards the selection of a codon, while the second summation penalizes the simultaneous selection of multiple synonymous codons within the same amino-acid group $a$. The parameters $\epsilon$ and $\mathrm{infty}$ jointly enforce the one-hot constraint, ensuring that exactly one synonymous codon is selected for each amino acid.
Combining all contributions yields the binary quadratic Hamiltonian

\begin{equation}
H(\mathbf{q})
=
\sum_i Q_{ii}q_i
+
\sum_{i<j}Q_{ij}q_iq_j,
\label{eq:QUBO}
\end{equation}

 where the diagonal coefficients contain contributions from codon usage, GC-content optimization, and the selection constraint, while the off-diagonal coefficients arise from GC-content regulation, repeated-nucleotide penalties, and one-hot constraint interactions. 
%For implementation on gate-based quantum computers, the binary variables are mapped onto qubit operators through $q_i=\frac{1-Z_i}{2}$, 
% % \begin{equation}
% % q_i=\frac{1-Z_i}{2},
% % \label{eq:mapping}
% % \end{equation}
% where $Z_i$ denotes the Pauli-$Z$ operator acting on qubit $i$. Substituting $q_i=\frac{1-Z_i}{2}$ into Eq.~(\ref{eq:QUBO}) yields the Pauli Hamiltonian

% \begin{equation}
% \hat{H}
% =
% E_0
% +
% \sum_i h_i Z_i
% +
% \sum_{i<j}J_{ij}Z_iZ_j,
% \label{eq:PauliHamiltonian}
% \end{equation}

% which serves as the cost Hamiltonian optimized by the VQE, SVQE, and QAOA algorithms throughout this work.

To implement the QUBO formulation on gate-based quantum computers, each binary variable is mapped to a qubit operator as $q_i=(1-Z_i)/2$, where $Z_i$ denotes the Pauli-$Z$ operator acting on qubit $i$. Substituting this mapping into Eq.(\ref{eq:QUBO}) yields the Ising Hamiltonian
\begin{equation}
\hat{H}
=
E_0
+
\sum_i h_i Z_i
+
\sum_{i<j}J_{ij}Z_iZ_j,
\label{eq:PauliHamiltonian}
\end{equation}
This Hamiltonian is used as the cost operator for the VQE, SVQE, and QAOA calculations performed in this work.

% ======================================================
\section{Results}
% ======================================================

\subsection{Three-Amino-Acid Model}
\begin{figure}[!t]
\centering
\subfloat[Energy spectrum.\label{fig:energy_spectrum}]{%
  \includegraphics[
    width=0.44\columnwidth,
    trim={2mm 0 2mm 0},
    clip
  ]{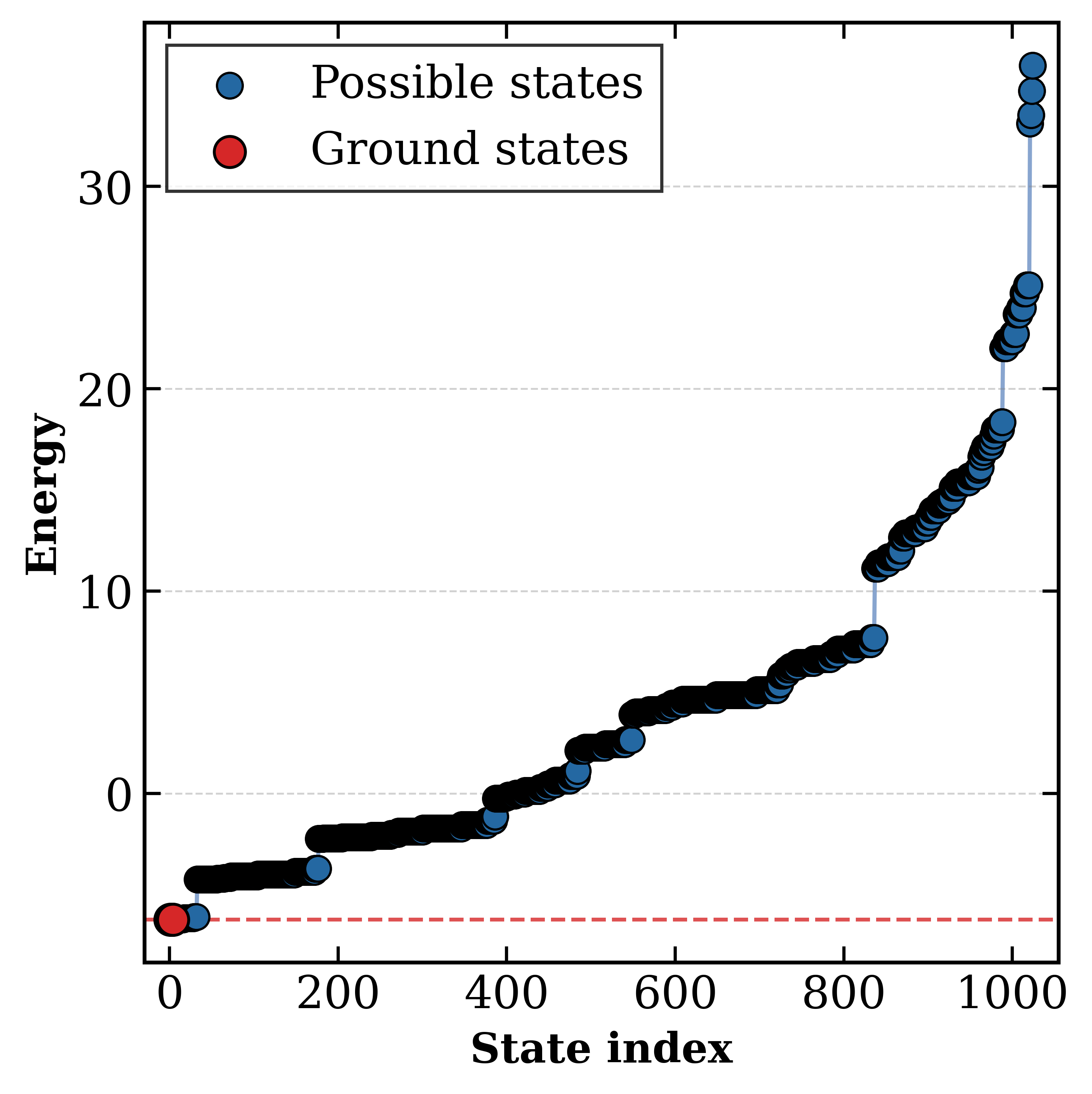}%
}%
\hspace{-0.001\columnwidth}%
\subfloat[Interaction graph.\label{fig:interaction_graph}]{%
    \raisebox{4.3mm}{%
  \includegraphics[
    width=0.5\columnwidth,
    trim={1mm 0 0.4mm 0},
    clip
  ]{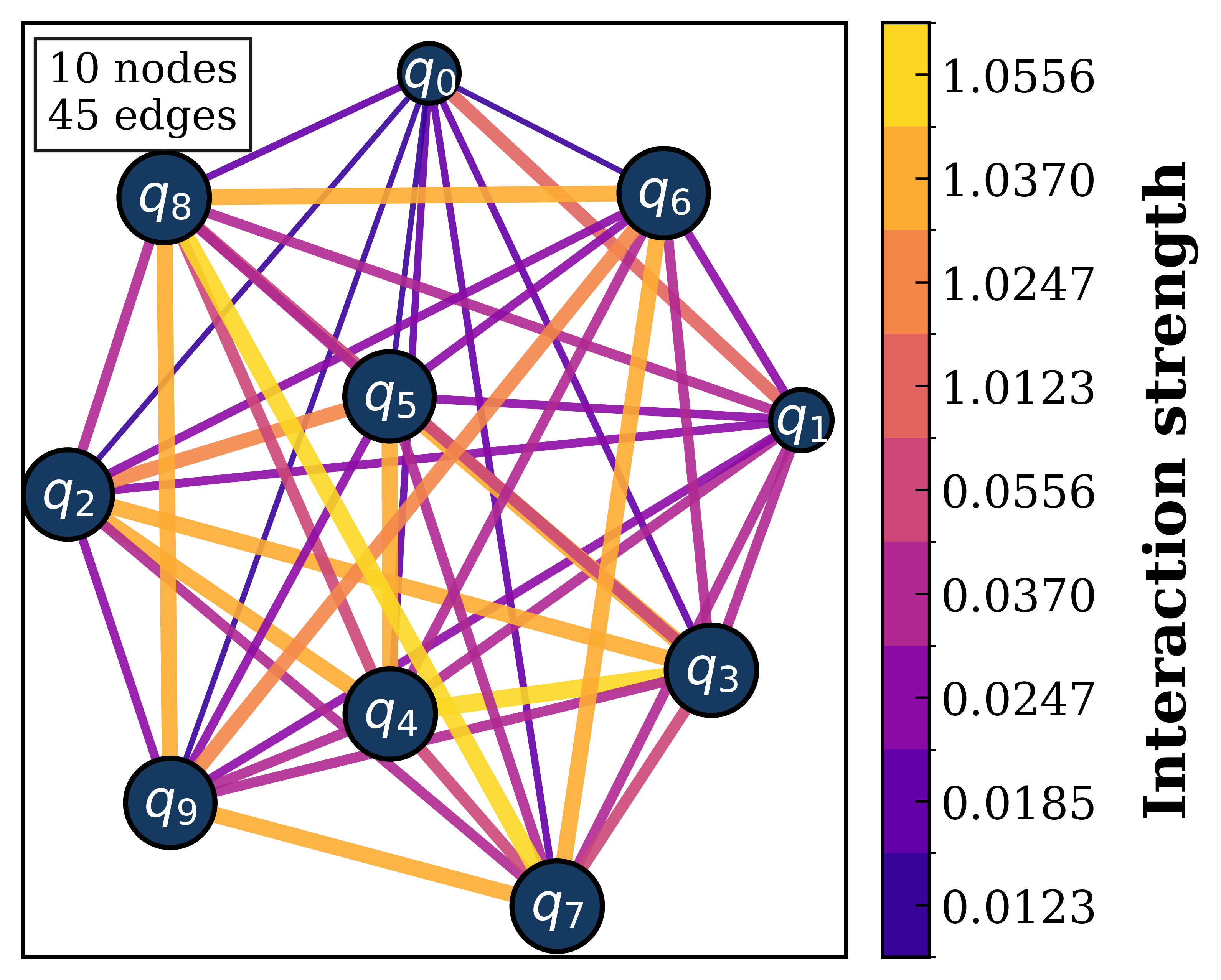}%
  }
}
\caption{Properties of the codon-optimization Hamiltonian:
    (a) energies of the 1024 possible states for Glu-Gly-Pro sequence, and
    (b) the corresponding interaction graph, where nodes represent
    binary variables and edge colors and widths indicate the interaction strengths for this case.}
\label{fig:ising_properties}
\vspace{-13pt}
\end{figure}

%\label{AA}
To validate the proposed framework, we first consider a benchmark codon optimization problem based on a three-amino acid sequence: Glutamic acid (Glu)--Glycine (Gly)--Proline (Pro). Glu has two synonymous codons (GAA and GAG), Gly has four (GGA, GGC, GGG, and GGT), and Pro has four (CCA, CCC, CCG, and CCT), resulting in ten candidate codons in total which also determine the length of bitstrings. Under the one-hot encoding scheme, each candidate codon is assigned to a single binary variable (qubit), yielding a 10-qubit ( $q_{1}$ to $q_{10}$) optimization problem. Qubits $q_{1}$ and $q_{2}$ represent the two synonymous codons of Glu, qubits $q_{3}$--$q_{6}$ represent the four synonymous codons of Gly, and qubits $q_{7}$--$q_{10}$ represent the four synonymous codons of Pro. Within each group, a one-hot constraint enforces that exactly one qubit takes the value $1$, thereby restricting the solution space to $32$ valid bitstrings of length $10$, each corresponding to a unique coding sequence for our model. Figure~\ref{fig:ising_properties} shows the energy spectrum of $2^{10}=1024$ possible bitstrings for this study, of which 32 satisfy the one-hot constraints, together with the corresponding interaction graph.
% \begin{figure}[!t]
%     \centering

% \subfloat[Energy spectrum.\label{fig:energy_spectrum}]{%
%     \includegraphics[
%         width=0.445\columnwidth
%     ]{codon_energy_spectrum.png}%
% }%
% \hspace{1mm}
% \hfill%
% \subfloat[Ising interaction graph.\label{fig:interaction_graph}]{%
%     \raisebox{3.5mm}{%
%         \includegraphics[
%             width=0.5\columnwidth
%         ]{ising_interaction_graph.png}%
%     }%
%     }

%     \caption{Properties of the codon-optimization Ising Hamiltonian:
%     (a) energies of the possible states ordered by increasing energy,
%     with the degenerate ground states highlighted; and
%     (b) the corresponding interaction graph, where nodes represent
%     binary variables and edge colors and widths indicate the
%     interaction strengths.}

%     \label{fig:hamiltonian_properties}
% \end{figure}
% To validate the proposed framework, we first consider a toy model consisting of the three-amino-acid sequence:
% Glutamic acid (Glu) - Glycine (Gly) - Proline (Pro).
% The corresponding synonymous codons are:
% \begin{align}
% \text{Glu:} &\quad \{\mathrm{GAA},\mathrm{GAG}\}, \nonumber\\
% \text{Gly:} &\quad \{\mathrm{GGA},\mathrm{GGC},\mathrm{GGG},\mathrm{GGU}\}, \nonumber\\
% \text{Pro:} &\quad \{\mathrm{CCA},\mathrm{CCC},\mathrm{CCG},\mathrm{CCU}\}. \nonumber
% \end{align}
%\subsection{Units}
% \begin{itemize}
% \item Use either SI (MKS) or CGS as primary units. (SI units are encouraged.) English units may be used as secondary units (in parentheses). An exception would be the use of English units as identifiers in trade, such as ``3.5-inch disk drive''.
% \item Avoid combining SI and CGS units, such as current in amperes and magnetic field in oersteds. This often leads to confusion because equations do not balance dimensionally. If you must use mixed units, clearly state the units for each quantity that you use in an equation.
% \item Do not mix complete spellings and abbreviations of units: ``Wb/m\textsuperscript{2}'' or ``webers per square meter'', not ``webers/m\textsuperscript{2}''. Spell out units when they appear in text: ``. . . a few henries'', not ``. . . a few H''.
% \item Use a zero before decimal points: ``0.25'', not ``.25''. Use ``cm\textsuperscript{3}'', not ``cc''.)
% \end{itemize}
% Thus, the optimization problem contains $32$ %$2 \times 4 \times 4 = 32$ 
% possible coding sequences, represented by ten binary variables under the one-hot encoding scheme.

For the three-amino-acid benchmark, the weight parameters were chosen as
$c_f=0$, $c_{\mathrm{GC}}=1$, $c_R=0$, $\rho_T=0.5$, $\epsilon=2$, and
$\mathrm{infty}=4$. The codon-frequency ($c_f$) and repeated-nucleotide ($c_R$) objectives were disabled because they are not meaningful for our short coding sequence example. The GC-content weight was set to unity, it simply activates the GC-content objective in Hamiltonian with unit weight. Consequently, the optimization is governed by the GC-content objective together with the one-hot codon-selection constraint.
The parameter $\rho_T=0.5$ specifies a target GC content of $50\%$, such that
candidate coding sequences with GC contents closer to this value are assigned
lower energies, while sequences with higher or lower GC contents remain
feasible but receive a larger penalty according to their deviation from the
target. The reason we assign a value to $\rho_T$ is that the objective of GC-content is not a maximization problem. In practical codon optimization, it is an optimization toward the desired GC content of the target host organism. The parameters $\epsilon=2$ and $\mathrm{infty}=4$ determine the contribution of the one-hot penalty term in the Hamiltonian. In the one-hot penalty term, the parameters $\epsilon$ and $\mathrm{infty}$ work together to enforce valid codon selection. The linear parameter $\epsilon$ favors the selection of a codon from each synonymous codon group, preventing the all-zero configuration from becoming energetically favorable. In contrast, the parameter $\mathrm{infty}$ penalizes the simultaneous selection of multiple codons within the same group. Together, these parameters ensure that configurations containing exactly one selected codon for each amino acid achieve the lowest constraint energy. For this parameter setting, preliminary exact diagonalization and classical optimization show that the Hamiltonian possesses multiple degenerate ground states. This benchmark is intentionally retained because it enables a more rigorous evaluation of the investigated algorithms based not only on their ability to identify the minimum energy, but also to recover the complete set of degenerate optimal coding sequences.

Table~\ref{tab:three_aa_simulator} summarizes the results obtained using exact diagonalization, classical optimization, quantum annealing, VQE, SVQE, and QAOA for the three-amino-acid benchmark. Exact diagonalization identifies four degenerate ground states with the same minimum energy of $-6.2469$, corresponding to four distinct coding sequences. Classical optimization, the D-Wave sampler, SVQE, and QAOA recover the complete set of four ground states. The probabilities and computational times reported in Table~I correspond to one representative run; repeated independent runs reproduced the same ground-state solutions in the same order, confirming the reproducibility of these results. Although the recovered energies and coding sequences are identical, the sampling probabilities differ among the D-Wave, SVQE, and QAOA methods because they generate different sampling distributions. In contrast, VQE exhibited strong dependence on the optimization run in the simulator tests and did not consistently recover the ground-state solutions. Owing to this lack of stability at the simulator level, VQE was not advanced to hardware implementation in this study, and no representative probability or hardware execution time is reported. 
%This agreement confirms that the proposed Pauli Hamiltonian faithfully preserves the original optimization problem.
\begin{table}[!t]
\setlength{\tabcolsep}{1.0pt}
\centering
\footnotesize
\caption{Simulator results for the three-amino-acid benchmark.
All methods recover the same four degenerate ground states
with identical minimum energy.}
\label{tab:three_aa_simulator}
\renewcommand{\arraystretch}{1.05}

\begin{tabular}{l c c l c c}
\hline
{\scriptsize\textbf{Method}} &
{\scriptsize\textbf{GS bitstring}} &
{\scriptsize\textbf{Energy}} &
{\scriptsize\textbf{DNA}} &
{\scriptsize\textbf{Prob.}} &
{\scriptsize\textbf{Time (s)}} \\
\hline

\multirow{4}{*}{\shortstack[l]{Exact\\diagonal.}}
&1000010001&$-6.2469$&GAAGGTCCT&--&\multirow{4}{*}{0.003439 }\\
&1000011000&$-6.2469$&GAAGGTCCA&--&\\
&1010001000&$-6.2469$&GAAGGACCA&--&\\
&1010000001&$-6.2469$&GAAGGACCT&--&\\
\hline

\multirow{4}{*}{\shortstack[l]{Classical\\optimization}}
&1000010001&$-6.2469$&GAAGGTCCT&--&\multirow{4}{*}{0.011104 }\\
&1000011000&$-6.2469$&GAAGGTCCA&--&\\
&1010000001&$-6.2469$&GAAGGACCT&--&\\
&1010001000&$-6.2469$&GAAGGACCA&--&\\
\hline

\multirow{4}{*}{\shortstack[l]{D-Wave\\sampler}}
&1010000001&$-6.2469$&GAAGGACCT&0.006500&\multirow{4}{*}{1.80697 }\\
&1000010001&$-6.2469$&GAAGGTCCT&0.006167	&\\
&1000011000&$-6.2469$&GAAGGTCCA&0.002833&\\
&1010001000&$-6.2469$&GAAGGACCA&0.004000	&\\
\hline

VQE
& \multicolumn{3}{c}{Run-dependent}
& --
& -- \\
\hline

\multirow{4}{*}{SVQE}
&1000010001&$-6.2469$&GAAGGTCCT&0.000346&\multirow{4}{*}{6.817 }\\
&1000011000&$-6.2469$&GAAGGTCCA&0.001564&\\
&1010001000&$-6.2469$&GAAGGACCA&0.000013&\\
&1010000001&$-6.2469$&GAAGGACCT&0.000128&\\
\hline

\multirow{4}{*}{QAOA}
&1000010001&$-6.2469$&GAAGGTCCT&0.000977&\multirow{4}{*}{131.587 }\\
&1000011000&$-6.2469$&GAAGGTCCA&0.000977&\\
&1010001000&$-6.2469$&GAAGGACCA&0.002930&\\
&1010000001&$-6.2469$&GAAGGACCT&0.001953&\\
\hline

\end{tabular}
\end{table}
Table~\ref{tab:svqe_hardware} summarizes representative SVQE results obtained on the \texttt{ibm\_berlin} quantum processor using a single initialization and sets of five and ten random initializations. The number of random initializations was increased to reduce sensitivity to the initial variational parameters and to provide broader exploration of the optimization landscape. Although increasing the number of random initializations increases the computational cost, it improves exploration of the variational parameter space, reduces sensitivity to unfavorable starting points, and increases the likelihood of obtaining the optimal solutions. The reported probabilities are the measured frequencies from one representative execution of each setting and are not averages over multiple executions. Repeated calculations produced the same qualitative probability distribution and ordering of the reported solutions. For the single-initialization case, the reported computational time is the total QPU usage required for that optimization. For the five- and ten-initialization cases, the reported time is the total QPU usage accumulated over all five or ten independent optimizations, respectively, including all optimizer-driven circuit evaluations but excluding queue waiting time. Thus, the timing values represent the complete computational cost of each initialization strategy.

\begin{table}[!t]
\setlength{\tabcolsep}{2.0pt}
\centering
\footnotesize
\renewcommand{\arraystretch}{1.05}

\caption{SVQE results obtained on the IBM Quantum backend \texttt{ibm\_berlin}.}
\label{tab:svqe_hardware}

\begin{tabular}{>{\centering\arraybackslash}p{1.65cm} l c l c c}
\hline
{\scriptsize\textbf{Runs}} &
{\scriptsize\textbf{Bitstring}} &
{\scriptsize\textbf{Energy}} &
{\scriptsize\textbf{DNA}} &
{\scriptsize\textbf{Prob.}} &
{\scriptsize\textbf{Time (s)}} \\
\hline

\multirow{5}{=}{\centering Single\\run}
&1000010001&$-6.2469$&GAAGGTCCT&0.00050&\multirow{5}{*}{1504.599 }\\
&1000011000&$-6.2469$&GAAGGTCCA&0.00025&\\
&1010000001&$-6.2469$&GAAGGACCT&0.00125&\\
&0100010001&$-6.2222$&GAGGGTCCT&0.00050&\\
&0100011000&$-6.2222$&GAGGGTCCA&0.00025&\\
\hline

\multirow{4}{=}{\centering 5 random\\init.}
&1000011000&$-6.2469$&GAAGGTCCA&0.00025&\multirow{4}{*}{8035.612 }\\
&0100010001&$-6.2222$&GAGGGTCCT&0.00025&\\
&0100011000&$-6.2222$&GAGGGTCCA&0.00050&\\
&1000101000&$-6.2222$&GAAGGGCCA&0.00125&\\
\hline

\multirow{4}{=}{\centering 10 random\\init.}
&1000010001&$-6.2469$&GAAGGTCCT&0.004000&\multirow{4}{*}{11614.559 }\\
&1000011000&$-6.2469$&GAAGGTCCA&0.00150&\\
&1010001000&$-6.2469$&GAAGGACCA&0.00925&\\
&1010000001&$-6.2469$&GAAGGACCT&0.00050&\\
\hline

\end{tabular}
\end{table}
Table~\ref{tab:qaoa_hardware} summarizes representative QAOA results obtained on the \texttt{ibm\_berlin} quantum processor under different hardware-execution settings. With the default configuration and 4096 measurement shots, QAOA identified three of the four degenerate ground states. Increasing the measurement count to 6000 or 10000 shots allowed all four ground states to be observed. The same result was obtained using dynamical decoupling with 4096 shots, whereas gate twirling did not produce the complete ground-state set under the examined conditions. Overall, the results demonstrate that QAOA sampling on noisy hardware is sensitive to the execution configuration and that increased measurement statistics or suitable error-suppression techniques can improve exploration of the degenerate solution space. The hardware execution time is shorter than the simulator time because the QAOA parameters were predefined for the hardware run, avoiding the repeated circuit evaluations required for variational parameter optimization.

\begin{table}[!t]
\setlength{\tabcolsep}{1.8pt}
\centering
\footnotesize
\renewcommand{\arraystretch}{1.15}
\caption{QAOA results obtained on the IBM Quantum backend \texttt{ibm\_berlin} under different hardware execution settings. DD and GT denote dynamical decoupling and gate twirling, respectively.}
\label{tab:qaoa_hardware}

\begin{tabular}{c c c c c c}
\hline
\textbf{Design} &
\textbf{Bitstrings} &
\textbf{Energies} &
\textbf{DNA Sequences} &
\textbf{Probability} &
\textbf{Time (s)} \\
\hline

\begin{tabular}[c]{@{}c@{}}
4096\\
shots
\end{tabular}
&
\begin{tabular}[c]{@{}c@{}}
1000010001\\
1000011000\\
1010001000\\
0100010001
\end{tabular}
&
\begin{tabular}[c]{@{}c@{}}
$-6.2469$\\
$-6.2469$\\
$-6.2469$\\
$-6.2222$
\end{tabular}
&
\begin{tabular}[c]{@{}c@{}}
GAAGGTCCT\\
GAAGGTCCA\\
GAAGGACCA\\
GAGGGTCCT
\end{tabular}
&
\begin{tabular}[c]{@{}c@{}}
0.000488\\
0.002197\\
0.000244\\
0.001709
\end{tabular}
&
\begin{tabular}[c]{@{}c@{}}
\\[-0.2em]
12.411 \\
\\
\end{tabular}
\\
\hline

\begin{tabular}[c]{@{}c@{}}
4096 shots\\
+ GT
\end{tabular}
&
\begin{tabular}[c]{@{}c@{}}
1000010001\\
1000011000\\
1010001000\\
0100010001
\end{tabular}
&
\begin{tabular}[c]{@{}c@{}}
$-6.2469$\\
$-6.2469$\\
$-6.2469$\\
$-6.2222$
\end{tabular}
&
\begin{tabular}[c]{@{}c@{}}
 GAAGGTCCT \\
GAAGGTCCA \\
 GAAGGACCA \\
 GAGGGTCCT \\
\end{tabular}
&
\begin{tabular}[c]{@{}c@{}}
0.000488\\
0.001709\\
0.001709\\
0.000732
\end{tabular}
&
\begin{tabular}[c]{@{}c@{}}
\\[-0.2em]
14.137 \\
\\
\end{tabular}
\\
\hline

6000 shots
&
\begin{tabular}[c]{@{}c@{}}
1000010001\\
1000011000\\
1010001000\\
1010000001
\end{tabular}
&
\begin{tabular}[c]{@{}c@{}}
$-6.2469$\\
$-6.2469$\\
$-6.2469$\\
$-6.2469$
\end{tabular}
&
\begin{tabular}[c]{@{}c@{}}
GAAGGTCCT\\
GAAGGTCCA\\
GAAGGACCA\\
GAAGGACCT
\end{tabular}
&
\begin{tabular}[c]{@{}c@{}}
0.000167\\
0.001000\\
0.001667\\
0.000500
\end{tabular}
&
\begin{tabular}[c]{@{}c@{}}
\\[-0.2em]
27.910 \\
\\
\end{tabular}
\\
\hline

\begin{tabular}[c]{@{}c@{}}
4096 shots\\
+ DD
\end{tabular}
&
\begin{tabular}[c]{@{}c@{}}
1000010001\\
1000011000\\
1010001000\\
1010000001
\end{tabular}
&
\begin{tabular}[c]{@{}c@{}}
$-6.2469$\\
$-6.2469$\\
$-6.2469$\\
$-6.2469$
\end{tabular}
&
\begin{tabular}[c]{@{}c@{}}
GAAGGTCCT\\
GAAGGTCCA\\
GAAGGACCA\\
GAAGGACCT
\end{tabular}
&
\begin{tabular}[c]{@{}c@{}}
0.000244\\
0.000732\\
0.002441\\
 0.000244
\end{tabular}
&
\begin{tabular}[c]{@{}c@{}}
\\[-0.2em]
25.809 \\
\\
\end{tabular}
\\
\hline

10000 shots
&
\begin{tabular}[c]{@{}c@{}}
1000010001\\
1000011000\\
1010001000\\
1010000001
\end{tabular}
&
\begin{tabular}[c]{@{}c@{}}
$-6.2469$\\
$-6.2469$\\
$-6.2469$\\
$-6.2469$
\end{tabular}
&
\begin{tabular}[c]{@{}c@{}}
GAAGGTCCT\\
GAAGGTCCA\\
GAAGGACCA\\
GAAGGACCT
\end{tabular}
&
\begin{tabular}[c]{@{}c@{}}
0.000800\\
0.000700\\
0.002100\\
0.000400
\end{tabular}
&
\begin{tabular}[c]{@{}c@{}}
\\[-0.2em]
17.726 \\
\\
\end{tabular}
\\
\hline

\end{tabular}
\end{table}

\section{Discussion}
The present work establishes a complete gate-based quantum computing framework for codon optimization, including Pauli-operator representation of Hamiltonian, implementation using different variational quantum algorithms, and validation on both quantum simulators and real superconducting quantum hardware. In first step, the three-amino-acid benchmark is employed as a proof-of-concept problem to systematically evaluate algorithm performance and identify hardware execution settings that maximize recovery of the degenerate ground-state manifold. The results show that execution parameters, including the number of measurement shots and error-suppression techniques such as dynamical decoupling, have an important impact on the solution. These observations provide practical guidelines for selecting algorithmic and hardware configurations in our future studies. Although the present work focuses on a small benchmark, the proposed methodology is intended as a scalable framework that will be extended to larger codon optimization problems. Preliminary investigations on longer amino acid sequences have been initiated and will be reported in future works, together with a comprehensive assessment of scalability and performance on next-generation quantum processors.
\section{Quantum Algorithm Implementation}

\subsubsection{Variational Quantum Eigensolver (VQE)}
The VQE implementation was performed using Qiskit (version 2.2.x). To identify an appropriate variational configuration, several ansatz circuits, classical optimizers, and circuit depths were evaluated. The final implementation employed a \texttt{TwoLocal} ansatz composed of single-qubit $R_y$ rotation gates and controlled-$Z$ (CZ) entangling gates with four repetitions ($\mathrm{reps}=4$). The variational parameters were optimized using the L-BFGS-B optimizer with a maximum of 30\,000 iterations. A statevector estimator was used for simulator calculations. This combination of ansatz, optimizer, and circuit depth consistently reproduced the exact ground-state energy.  The measured bitstrings were decoded into DNA sequences, and only valid one-hot encoded solutions were retained for analysis.

\subsubsection{Sampling Variational Quantum Eigensolver (SVQE)}

SVQE simulations were performed using the Qiskit implementation of the Sampling Variational Quantum Eigensolver. Following an evaluation of different ansatz configurations, circuit depths, and optimization settings, the final implementation employed a \texttt{TwoLocal} ansatz with $R_y$ rotation gates, controlled-$Z$ (CZ) entangling gates, and two repetitions ($\mathrm{reps}=2$). The variational parameters were optimized using the L-BFGS-B optimizer with a maximum of 300 iterations, and ideal quantum measurements were simulated using a \texttt{StatevectorSampler}. The sampled bitstrings were filtered by the one-hot constraint, decoded into DNA sequences, evaluated using their Ising energies, and ranked by energy. Their sampling probabilities were reported separately to quantify the occurrence frequencies of the optimal sequences and characterize the degenerate ground-state manifold.

\subsubsection{SVQE on Real Quantum Hardware}

The SVQE hardware implementation was performed on the IBM Quantum backend \texttt{ibm\_berlin} using the IBM Runtime service. The same Pauli Hamiltonian and \texttt{TwoLocal} ansatz as in the simulator were employed. The parameterized circuit was transpiled using the Qiskit preset pass manager (optimization level 1), and the Hamiltonian was transformed according to the transpiled layout before execution. Circuit parameters were optimized using the Runtime \texttt{Sampler} primitive and the COBYLA optimizer (maximum 50 iterations), with 4000 measurement shots per circuit evaluation unless otherwise stated.
To improve sampling diversity and reduce sensitivity to the initial variational parameters, the hardware optimization was additionally repeated using five and ten independent random parameter initializations, with initial parameters sampled uniformly from $[0,2\pi)$ while all other settings remained unchanged. The measured bitstrings were post-processed using the same procedure as in the simulator. %The valid solutions from all runs were combined, retaining only the highest-probability occurrence of duplicated bitstrings, and were finally ranked by increasing energy and, for degenerate states, decreasing sampling probability. This strategy improves the likelihood of recovering the complete degenerate ground-state manifold on noisy quantum hardware.

\subsubsection{Quantum Approximate Optimization Algorithm (QAOA)}

QAOA simulations were performed using the Qiskit implementation of the Quantum Approximate Optimization Algorithm with the Pauli-encoded codon optimization Hamiltonian as the cost Hamiltonian and a \texttt{StatevectorSampler} for ideal measurements. A single alternating layer ($p=1$) and the COBYLA optimizer (maximum 300 iterations) were employed. After optimization, the sampled bitstrings were post-processed by enforcing the one-hot constraint, decoding valid solutions into DNA sequences, evaluating their corresponding Ising energies, and ranking them by increasing energy.
The hardware implementation was carried out on the IBM Quantum backend \texttt{ibm\_berlin} using a single-layer ($p=1$) QAOA Ansatz constructed from the Pauli cost Hamiltonian. The circuit was transpiled with the Qiskit preset pass manager (optimization level 1), and the Hamiltonian was mapped onto the transpiled layout before execution with the IBM Runtime \texttt{SamplerV2} primitive. Fixed parameters ($\beta=0.5$, $\gamma=0.8$) were employed, while different hardware execution settings, including increased measurement shots, dynamical decoupling, and gate twirling, were investigated. The measured bitstrings were post-processed following the same procedure as in the simulator.

\subsubsection{Quantum Annealing Benchmark}

The QUBO formulation was sampled on a D-Wave quantum annealer using the Ocean SDK. Minor embedding was performed automatically using the \texttt{EmbeddingComposite}, with a chain strength of 30 and 6000 reads.
% ======================================================
\section{Conclusions}
% ======================================================
This work presents an initial investigation of gate-based quantum algorithms for the codon optimization problem through an Ising Hamiltonian formulation suitable for quantum computation. A computational workflow based on VQE, SVQE, and QAOA was developed and benchmarked against exact diagonalization, classical optimization, and quantum annealing. The current study focuses on a three-amino-acid benchmark, for which all investigated methods successfully recovered the complete set of degenerate ground-state solutions identified by exact diagonalization. In addition, hardware experiments on IBM Quantum processors were performed to evaluate the practical implementation of SVQE and QAOA under different execution settings. These experiments provide insight into the influence of hardware parameters, such as measurement shots and error-suppression techniques, on solution quality and ground-state recovery. The results presented in this work correspond to optimized algorithmic configurations obtained through systematic evaluation of circuit ansätze, optimization strategies, initialization schemes, and hardware execution parameters. The selected configurations were chosen because they provide reproducible performance and the closest agreement with exact solutions while accounting for the noise and hardware limitations of current quantum quantum processors.

As a work in progress, the primary contribution of this study is the establishment and validation of a gate-based quantum workflow for codon optimization. The implementation developed in this work provides a flexible framework for systematically investigating larger codon optimization problems as quantum hardware and algorithms continue to mature, with more comprehensive results to be reported in future work. %Ongoing work is extending the proposed methodology to longer amino-acid sequences, refining algorithmic parameters, and exploring improved quantum hardware configurations and error-mitigation strategies to further enhance performance and scalability.

\end{document}